\documentclass[a4paper,preprint,showkeys]{amsproc}
\usepackage{mathrsfs}
\usepackage{amsmath,booktabs}
\usepackage[T4, OT1]{fontenc}
\usepackage[arrow, matrix, curve]{xy}
\usepackage{newunicodechar}
\usepackage{graphicx}
\usepackage{bm}
\usepackage[dvips]{epsfig}
\usepackage{rotating}
\usepackage{amsmath,amssymb,amsfonts,amsthm}
\usepackage{amsmath,amscd}
\usepackage{subfig}
\usepackage{caption}
\usepackage{pst-all}
\usepackage{dcolumn}
\usepackage[thinc]{esdiff}
\usepackage{amsthm}
\usepackage{dsfont}
\usepackage{amssymb}
\usepackage[hyphens]{url} 
\usepackage[dvips]{graphicx}
\usepackage{xparse}
\usepackage{physics}
\usepackage{mathrsfs}
\makeatletter
\@namedef{subjclassname@2024}{\textup{2024} Mathematics Subject Classification}
\makeatother
\theoremstyle{plain}

\theoremstyle{definition}

\theoremstyle{remark}
 
 \numberwithin{equation}{section}

\newcommand{\eg}{e.\,g.\ }
\newcommand{\ie}{i.\,e.\ }

\title[Riemann $\zeta$-Regularisation of Path Integrals.]{Riemann-$\zeta$-Regularisation of \\ Feynman Path Integrals.}
\author[Belardinelli]{Cyril Belardinelli} 
\address{ 
Kantonsschule Solothurn\\
Solothurn\\
Switzerland}
\email{cyril.belardinelli@ksso.ch}
\begin{document}
\vspace{18mm} \setcounter{page}{1} \thispagestyle{empty}
\begin{abstract}
The Feynman propagator of a charged particle confined to an anisotropic harmonic oscillator potential 
and moving in a crossed electro-magnetic field is calculated in a conceptually new way. The calculation is based on the expansion of the path variable into a complex Fourier series. The path integral then becomes an infinite product of Gaussian integrals. This product is divergent. It turns out that we can regularize this product by using the  $\zeta$-function. It is a remarkable fact that the $\zeta$- function is so well suited as a regularizator for divergent path integrals.
\end{abstract}
\maketitle
\section{\label{intro}Introduction}
In the present paper we recalculate the Feynman propagator of a charged particle submitted to both an anisotropic in-plane oscillator and a constant magnetic field aligned along the $z$-axis, \ie $\vb{B}=(0,0,B)$. In addition, a constant electric field perpendicular to the magnetic field is present, $\vb{E}=(0,E_x, E_y)$.\\ 
The standard procedure of calculating a path integral is by dividing up the time interval $[0,T]$ in N time slices. In the limit $N \to \infty$, the integral over positions at each time slice can be said to be an integral over all possible continuous paths. 
However, this method has several conceptional drawbacks; the continuous paths over which one integrates are almost certainly nowhere differentiable. This means that kinetic energy is a diverging quantity with probability one. An additional drawback is the occurrence of an infinite Normalization constant in the limit of infinitely many time slices.
Furthermore, the evaluation of path integrals for Lagrangians with a velocity dependent potential, such as the magnetic interaction by the method of time slicing is more delicate; it leads to ambiguities when not correctly discretized. A correct calculation requires the vector potential $\vb{A}$ be evaluated at the midpoint of each straight segment of the path\cite{schulman:1981}.\\
\\
In the present paper we use a conceptually new method which is based on regularizing the divergent path integral by the Riemann $\zeta$-function. Due to the two-dimensional nature of the problem we have to sum over all paths $\gamma$ in the plane running between an initial and a final point. We expand the deviation from the classical path $\eta:=\gamma-\gamma_{\text{cl}}$ in a complex Fourier series. 
However, this calculation leads to an infinite product (of one-dimensional integrals) which converges to zero. Remarkably, we can turn this non-sensical result into a meaningful one by $\zeta$-regularizing the infinite product.\\
Even though the Action is quadratic, a full calculation of both the Probability Amplitude and the classical Action is rather involved. The same problem can also be tackled by other methods; a possibility is the use of the well-known Van Vleck- Pauli- Morette (VVPM) formula which becomes exact in the case of quadratic Lagrangians \cite{cervero:2017}.\\
Another less known possibility is the use of the Gelfand-Yaglom formula (GY)\cite{gelfand:1960}.
The spirit of this paper is driven by the desire to give a more intelligibel meaning to Integration in infinite dimensional spaces which is at the core of path integration. Some parts may be called \textit{heuristic} because we \textquotedblleft blindly\textquotedblright  interchange integration variables without further justification.
\\
\section{\label{lagrange}Lagrangian}
The Lagrangian in consideration is given by;
\begin{equation*}
\mathcal{L}={\frac{m}{2}\qty(\dot{x}^2+\dot{y}^2)}-{\frac{m\omega_{x}^2}{2} x^2-\frac{m\omega_{y}^2}{2} y^2}
+q\vb{v}\vdot\vb{A}+q\vb{E}\vdot\vb{r}
\end{equation*}
In the symmetric gauge of the vector potential $\vb{A}$ the Lagrangian $\mathcal{L}$ reads as follows;
\begin{equation*}
\mathcal{L}={\frac{m}{2}\qty(\dot{x}^2+\dot{y}^2)}-{\frac{m\omega_{x}^2}{2} x^2-\frac{m\omega_{y}^2}{2} y^2}
+m\omega_{L}\qty(x\dot{y}-y\dot{x})+q\vb{E}\vdot\vb{r}
\end{equation*}
where $\omega_{L}=qB/2m$ is the Larmor frequency and $\omega_{x},\omega_{y}$ are the oscillator frequencies in each direction of the plane.\\
For the calculation of the Propagator it turns out to be convenient to express the Lagrangian in terms of complex coordinates;
\begin{eqnarray*}
&\gamma&:=x+iy \\ 
&\epsilon&:=E_x+iE_y
\end{eqnarray*}
where the function $\gamma(t):[0,T]\rightarrow\mathbb{C}$ is an element of the infinite dimensional Hilbert space $\qty(\mathcal{H},\expval{\,,\,})$ of square-integrable complex-valued functions defined on the time intervall $[0,T]$. The space $\mathcal{H}=L^2_{\mathbb{C}}[0,T]$ is endowed with the canonical inner product
given as;
\begin{equation*}
\expval{f|g}:=\frac{1}{T}\int\limits_0^T f(t)g(t)^{\ast}d{t}
\end{equation*}
In terms of the path variable $\gamma(t)$ the Lagrangian reads;
\begin{eqnarray*}
\mathcal{L}&=&\frac{m}{2}\dot{\gamma}\dot{\gamma}^{\ast}-\frac{m}{4}\qty(\omega_{x}^{2}+\omega_{y}^{2})\gamma\gamma^{\ast}+\frac{m}{8}\qty(\omega_{y}^{2}-\omega_{x}^{2})(\gamma^{2}+\text{c.c.})\\
&-&
\frac{im\omega_{L}}{2}(\gamma^{\ast}\dot{\gamma}-\gamma\dot{\gamma}^{\ast})+\frac{q}{2}{\epsilon}\gamma^{\ast}+\frac{q}{2}{\epsilon^{\ast}}\gamma
\end{eqnarray*}
The Action $S=\int \mathcal{L}d{t}$ is;
\begin{eqnarray*}
\label{action}
S[\gamma]&=&\frac{mT}{2}\expval{\dot{\gamma}|\dot{\gamma}}-\frac{mT}{4}\qty(\omega_{y}^{2}+\omega_{x}^{2}) \expval{\gamma|\gamma}+\frac{mT}{8}\qty(\omega_{y}^{2}-\omega_{x}^{2})\qty(\expval{\gamma|\gamma^{\ast}}+\text{c.c.})\\
&-&\frac{im\omega_{L}T}{2}\qty(\expval{\dot{\gamma}|{\gamma}}-\expval{\gamma|\dot{\gamma}})+\frac{qT}{2}\expval{\epsilon|\gamma}+\frac{qT}{2}\expval{\gamma|\epsilon}
\end{eqnarray*}
The Propagator then reads;
\begin{equation*}
\bra{\textbf{x}_{f}, T}\ket{\textbf{x}_{i},0} =\int\limits_{\gamma(0)}^{\gamma(T)}\mathcal{D}[\gamma(t)]\,e^{i S[\gamma]/\hbar}
\end{equation*}
where 
\begin{equation*}
\label{boundary}
\begin{split}
\textbf{x}_{i}&\equiv \gamma(0)\equiv x_0+iy_0\\
\textbf{x}_{f}&\equiv \gamma(T)\equiv x_1+iy_1
\end{split}
\end{equation*}
We presume the functional measure $\mathcal{D}[\gamma(t)]$ to be invariant under translation in the functional space $ \mathcal{H}=L^2_{\mathbb{C}}[0,T]$.
The measure can therefore be assumed to be invariant under the constant shift $\gamma_{cl} \in \mathcal{H}$ given by;
\begin{equation*}
\label{variable_trans}
\gamma(t) = \gamma_{\text{cl}}(t)+\eta(t)
\end{equation*}
The quantum fluctuation $\eta(t)$ describes a closed loop in the complex plane because the initial point and final point coincide, \ie $\eta(0)=\eta(T)=0$.
\\
\\
In terms of $\eta$ 
\begin{eqnarray*}
\bra{\textbf{x}_{f}, T}\ket{\textbf{x}_{i},0} =e^{i S_{\text{c}}/\hbar} \int\limits_{\eta(0)=0}^{\eta(T)=0}\mathcal{D}[\eta(t)]\,e^{i S[\eta]/\hbar}
\end{eqnarray*}
where $S_{\text{cl}}$ denotes the classical Action. The integral part in the latter expression is the so-called Probability Amplitude $\mathcal{A}$.
\section{\label{propagator}Probability Amplitude $\mathcal{A}$}
\subsection{Expanding the paths in a Fourier Series.}
In  the present section we calculate the Amplitude of the Propagator by expanding the function $\eta(t)$ in a complex Fourier series;
\begin{equation*}
\label{fourier_exp}
\eta(t)=\sum_{n=-\infty}^{+\infty}c_{n}\,e^{\frac{2\pi i n}{T}\cdot t}
\end{equation*} 
Plugging this into the Action $S[\eta]$ leads to:
\begin{equation*}
\begin{split}
 S/\hbar=qT/2\hbar\,\qty(\epsilon c_{0}^{\ast}+\epsilon^{\ast}c_{0})+\sum_{n=-\infty}^{+\infty}\qty(\alpha n^2+\beta n-\lambda)\abs{c_n}^2+\Delta\qty(c_{n}c_{-n}+c_{n}^{\ast}c_{-n}^{\ast})
\end{split}
\end{equation*}
where
\begin{eqnarray*}
\alpha&:=&\frac{2m\pi^2}{\hbar T} \quad \quad \quad \quad \,\,\,\,\beta:=\frac{2m\pi\omega_{L}}{\hbar}\\
\\
\lambda&:=&\frac{mT}{4\hbar}(\omega_{y}^2+\omega_{x}^2) \quad \Delta:=\frac{mT}{8\hbar}(\omega_{y}^2-\omega_{x}^2)
\end{eqnarray*}
\\
The Amplitude $\mathcal{A}$ can formally be written as follows\footnote{This is formally an infinite-dimensional integral. At this point a precise definition is lacking. Heuristically, we treat it in the same way as an integral in finite dimensions.}
\begin{equation}
\label{functional_integral}
\mathcal{A}=\int\limits_{\mathbb{C}^{\infty}}\mathrm{d}^{\infty}\,e^{iS/\hbar}\,\delta\qty(\sum_{n=-\infty}^{+\infty}c_{n})
\end{equation}
with short notation;
\begin{eqnarray*}
\mathrm{d}^{\infty}:=\prod_{n=-\infty}^{\infty}\mathrm{d}c_{n}\quad\text{and}\quad
\mathbb{C}^{\infty}:=\mathbb{C}\cross \mathbb{C}\cross \dots
\end{eqnarray*}
The Dirac-delta distribution $\delta(\cdot)$ must be included in the path integral \ref{functional_integral} in order to fulfill the boundary conditions $\eta(0)=\eta(T)=0$. The integration over the complex Fourier component $c_n$ is to be understood in the following sense;
\begin{equation*} 
\int_{\mathbb{C}} \mathrm{d}c_{n}:=\int_{\mathbb{R}^2} \mathrm{d}x_{n} \mathrm{d}y_{n}
\end{equation*}
\\
By splitting the real and imaginary part of $c_{n}:=x_{n}+iy_{n}$ we can cast the Action in the form;
\begin{equation*}
\begin{split}
 S/\hbar=C\qty(x_{0},y_{0})
+&\sum_{n=1}^{\infty}A_{n}\qty(x_{n}^2+y_{n}^2)+A_{-n}\qty(x_{-n}^2+y_{-n}^2)\\
+4\Delta&\sum_{n=1}^{\infty}x_{n}x_{-n}-y_{n}y_{-n}
\end{split}
\end{equation*}
where \\
\begin{eqnarray*}
C\qty(x_{0},y_{0})&:=&\qty(2\Delta-\lambda)x_{0}^2+qT E_{x}x_0/\hbar-\qty(2\Delta+\lambda)y_{0}^2+qT E_{y}y_0/\hbar\\
A_{n}&:=&\alpha n^2 +\beta n-\lambda
\end{eqnarray*}
\\
We write the Action $S$ as the sum of an $x$-dependent and $y$-dependent part, \ie
\begin{equation*}
S=S_{x}+S_{y}
\end{equation*}
where;
\begin{eqnarray*}
S_x/\hbar&=&ax_{0}+(2\Delta-\lambda)x_{0}^2+\sum_{n=1}^{+\infty}A_{n}x_{n}^2+A_{-n}x_{-n}^2+4\Delta x_{n}x_{-n}\\
S_y/\hbar&=&by_{0}-(2\Delta+\lambda)y_{0}^2+\sum_{n=1}^{+\infty}A_{n}y_{n}^2+A_{-n}y_{-n}^2-4\Delta y_{n}y_{-n}
\end{eqnarray*}
The delta function introduced in Eq.~\ref{functional_integral} can be separated as well. We can therefore factorize the Amplitude $\mathcal{A}$;
\begin{equation*}
\mathcal{A}=\mathcal{A}_{x}\mathcal{A}_{y}
\end{equation*}
where
\begin{eqnarray*}
\label{amplitudes}
\mathcal{A}_{x}&=&\int\limits_{\mathbb{R}^{\infty}}\prod_{n=-\infty}^{\infty}\mathrm{d}x_{n}\,e^{iS_{x}/\hbar}\,\delta\qty(\sum_{n=-\infty}^{+\infty}x_{n})\\
\mathcal{A}_{y}&=&\int\limits_{\mathbb{R}^{\infty}}\prod_{n=-\infty}^{\infty}\mathrm{d}x_{n}\,e^{iS_{y}/\hbar}\,\delta\qty(\sum_{n=-\infty}^{+\infty}y_{n})
\end{eqnarray*}
Next, we substitute the delta function $\delta(\cdot)$ by its integral representation;
\begin{eqnarray*}
\delta\qty(\sum_{n=-\infty}^{+\infty} x_{n})=\frac{1}{2\pi}\int\limits_{\mathbb{R}}\mathrm{d}k\exp({ik\sum_{n=-\infty}^{+\infty} x_{n}})\quad \text{and}\quad (x_{n}\leftrightarrow y_{n})
\end{eqnarray*}
This leads to;
\begin{equation*}
\mathcal{A}_x=\frac{1}{2\pi}\int_{\mathbb{R}} \mathrm{d}k \,I(k) \prod_{n=1}^{\infty}\,\, \int\limits_{\mathbb{R}^2}\mathrm{d}x_{n}\mathrm{d}x_{-n}\,e^{i\qty(A_{n}x_{n}^2+A_{-n}x_{-n}^2+4\Delta x_{n}x_{-n}+kx_{n}+kx_{-n})}
\end{equation*}
The integral $I(k)$ is given by the expression;
\begin{equation*}
I(k)=\int_{\mathbb{R}} \mathrm{d}x_{0}\,e^{i(2\Delta-\lambda)x_{0}^2+i(a+k)x_{0}}
=e^{-i\frac{(a+k)^2}{8\Delta-4\lambda}}\sqrt{\frac{i\pi}{2\Delta -\lambda}}
\end{equation*}
where;
\begin{equation*}
a:=\frac{q TE_x}{\hbar}\quad b:=\frac{qT E_y}{\hbar}
\end{equation*}
By taking advantage of matrix notation we can write in compact form;
\begin{equation}
\label{amplitude_simple}
\mathcal{A}_x=\frac{1}{2\pi}\int_{\mathbb{R}} \mathrm{d}k\,I(k) \prod_{n=1}^{\infty}\,\, \int\limits_{\mathbb{R}^2}\mathrm{d}^{2}X_{n}\,\exp\qty(i\, X_{n}^{\mathrm{t}}\,M_n X_{n}+i\, J^{\mathrm{t}}X_{n})
\end{equation}
where
\begin{equation*}
X_{n}^{t}:=\qty(x_{n},x_{-n}),\quad J^{t}=(k, k)
\end{equation*}
and
\begin{equation*}
M_{n}=\mqty(A_{n} & 2\Delta \\ 2 \Delta & A_{-n})
\end{equation*}\\
We note that the Amplitude $\mathcal{A}_{y}$ can be derived from $\mathcal{A}_{x}$ by substituting $\Delta$ with $-\Delta$ and $a$ with $b$.
The second integral in Eq.~\ref{amplitude_simple} is Gaussian. Its evaluation is;
\begin{equation*}
\int\limits_{\mathbb{R}^2}\mathrm{d}^{2}X_{n}\,\exp\qty(i\, X_{n}^{\mathrm{t}}\,M_n X_{n}+i\, J^{\mathrm{t}}X_{n})=\frac{i\pi}{\sqrt{\det M_n}}e^{-\frac{i}{4}J^{t}M_{n}^{-1}J}
\end{equation*}
Summa Summarum we have the following expressions for the Amplitudes $\mathcal{A}_x$,$\mathcal{A}_y$;
\begin{eqnarray*}
\mathcal{A}_x&=&\frac{1}{2\pi}\int_{\mathbb{R}} \mathrm{d}k\,e^{-i\frac{(a+k)^2}{8\Delta-4\lambda}}\sqrt{\frac{i\pi}{2\Delta -\lambda}}\prod_{n=1}^{\infty}\,\frac{i\pi}{\sqrt{\det M_n}}e^{-\frac{i}{4}J^{t}M_{n}^{-1}J}\\
\mathcal{A}_y&=&\frac{1}{2\pi}\int_{\mathbb{R}} \mathrm{d}k\,e^{-i\frac{(b+k)^2}{-8\Delta-4\lambda}}\sqrt{\frac{i\pi}{-2\Delta -\lambda}}\prod_{n=1}^{\infty}\,\frac{i\pi}{\sqrt{\det N_n}}e^{-\frac{i}{4}J^{t}N_{n}^{-1}J}
\end{eqnarray*}
where 
\begin{equation*}
N_{n}=\mqty(A_{n} & -2\Delta \\ -2 \Delta & A_{-n})
\end{equation*}\\
and
\begin{eqnarray*}
\det M_n&=&\det N_n=\qty(A_{n}A_{-n}-4\Delta^2)\\
&=&(\alpha n^2-\lambda)^{2}-\beta^{2} n^2-4\Delta^2
\end{eqnarray*}
\subsection{Evaluation of infinite products}
In the next step we have to evaluate the infinite product arising in the 
Amplitude $\mathcal{A}_x$ ($\mathcal{A}_y$);
\begin{equation*}
\label{first}
\prod_{n=1}^{\infty}\frac{i\pi}{\sqrt{\det M_n }}\exp(-\frac{i}{4}J^{T}M_n^{-1}J)
\end{equation*}
We split the product in two parts;
\begin{eqnarray*}
\Pi_1:&=&\prod_{n=1}^{\infty}\frac{i\pi}{\sqrt{\det M_n}}=\prod_{n=1}^{\infty}\frac{i\pi}{\sqrt{\qty(\alpha n^2-\lambda)^{2}-\beta^{2} n^2-4\Delta^2}}\\
\Pi_2:&=&\prod_{n=1}^{\infty}\exp(-\frac{i}{4}J^{T}M_n^{-1}J)
\end{eqnarray*}
In the first product $\Pi_1$ we factorize the denominator;
\begin{equation}
\sqrt{\det M_n}=\frac{i\pi}{\alpha n^2}\cdot\frac{1}{\sqrt{1-\omega_{+}^2/n^2}}\cdot\frac{1}{\sqrt{1-\omega_{-}^2/n^2}}
\end{equation}
where
\begin{eqnarray*}
\omega_{\pm}&=&\frac{\Omega_{+}\pm\Omega_{-}}{2}\cdot\qty(\frac{T}{2\pi})\\
\Omega_{\pm}&=&\sqrt{\omega_{c}^2+\qty(\omega_{x}\pm \omega_{y})^2}
\end{eqnarray*}
\\
Note that 
$\omega_{c}=\omega_{L}/2=qB/m$ denotes the cyclotron frequency.\\
\\
Applying the well-known Euler sine product gives;
\begin{equation}
\label{prod}
\Pi_1=\sqrt{\frac{\pi \omega_{+}}{\sin{\pi \omega_{+}}}}\cdot \sqrt{\frac{\pi \omega_{-}}{\sin{\pi \omega_{-}}}}\prod_{n=1}^{\infty}\frac{i \pi}{\alpha n^2}
\end{equation}
At this stage we encounter the first true obstacle which is unsurmountable by classical methods; the infinite product in Eq.~\ref{prod} converges to zero which evidently doesn't serve our purpose.\\We interpret this as a manifestation of the fact that the vast majority of paths (in the sense of some measure such as \eg the Wiener measure on the space of continuous functions, \ie Brownian motion) on the space $L^2_{\mathbb{C}}[0,T]$ are non-differentiable. 
This means that the Amplitudes for practically all paths close to the classical trajectory are wildly oscillating despite the fact that it is a stationary \textquotedblleft point\textquotedblright. These paths cancel each other out so that the entire \textquotedblleft sum of histories\textquotedblright  adds up to zero in every $\epsilon$-environment about the classical path. For this reason, we must somehow filter out these paths by an appropriate method. It turns out that Riemann's $\zeta$- function fullfills this task at perfection. It is an interesting question why the $\zeta$-function is so well suited for this task. We aren't able to answer this question in the present paper. That would be an interesting problem for further investigation.\\
\\
The $\zeta$-regularized product in \ref{prod} can be written as follows;
\begin{equation}
\qty(\log{\prod_{n=1}^{\infty}\frac{i\pi}{\alpha n^2}})_{\zeta-\text{reg}}=\log{\frac{i\pi}{\alpha}}\qty(\sum_{n=1}^{\infty}1)_{\zeta-\text{reg}}-2\qty(\sum_{n=1}^{\infty}\log{n})_{\zeta-\text{reg}}
\end{equation}
The $\zeta$-regularized sums are given by;
\begin{eqnarray*}
&&\zeta(0)=:\qty(\sum_{n=1}^{\infty}1)_{\zeta-\text{reg}}=-\frac{1}{2}\\
\\
&&\zeta^{\prime}(0)=:-\qty(\sum_{n=1}^{\infty}\log{n})_{\zeta-\text{reg}}=-\frac{1}{2}\log{2\pi}
\end{eqnarray*}
We have therefore the $\zeta-$regularized product;
\begin{equation*}
\qty(\prod_{n=1}^{\infty}\frac{i\pi}{\alpha n^2})_{\zeta-\text{reg}}=\frac{1}{2\pi}\sqrt{\frac{\alpha}{i \pi}}
\end{equation*}
This gives us the first product $\Pi_1$;
\begin{equation*}
\Pi_1=\frac{1}{2}\sqrt{\frac{\alpha\omega_{+}\omega_{-}}{i\pi\sin{\pi\omega_{+}}\sin{\pi\omega_{-}}}}
\end{equation*}
\\
The second product $\Pi_2$ remains to be calculated. It reads;
\\
\begin{equation}
\label{sum}
\Pi_2=\prod_{n=1}^{\infty}\exp(-\frac{i}{4}J^{T}M_n^{-1}J)=\exp(-\frac{i}{4}\sum_{n=1}^{\infty}J^{T}M_n^{-1}J)
\end{equation}
where
\begin{equation*}
\sum_{n=1}^{\infty}J^{T}M_n^{-1}J=\frac{2k^2}{\alpha}\cdot \underbrace{\sum_{n=1}^{\infty}\frac{n^2-A_x}{n^4-Bn^2+C}}_{:=D_{x}}
\end{equation*}
with parameters;
\begin{eqnarray*}
A_x&=&\frac{\lambda}{\alpha}+ 2\frac{\Delta}{\alpha}, \quad
A_y=\frac{\lambda}{\alpha}- 2\frac{\Delta}{\alpha}\\
B&=&\frac{\beta^2}{\alpha^2}+ 2\frac{\lambda}{\alpha}, \quad
C=\frac{\lambda^2}{\alpha^2}- 4\frac{\Delta^2}{\alpha^2}
\end{eqnarray*}
in terms of $\omega_{x}, \omega_{y}$ it reads;
\begin{eqnarray*}
A_x&=&\qty(\frac{T}{2\pi})^{2}\omega_{y}^2, \quad A_y=\qty(\frac{T}{2\pi})^{2}\omega_{x}^2 \\
\quad B&=&\qty(\frac{T}{2\pi})^{2}\qty(4\omega_{L}^2+\omega_{x}^2+\omega_{y}^2)\\
C&=&A_xA_y=\qty(\frac{T}{2\pi})^{4}\omega_{x}^2\omega_{y}^2
\end{eqnarray*}
\\
The sum $D_x$ can be calculated by a partial fraction expansion. A straightforward but somewhat lengthy calculation results in;
\\
\begin{eqnarray*}
D_{x}=
\frac{A_x}{2\omega_{+}^{2}\omega_{-}^2}-\pi A_x\frac{\omega_{+}\cot{\pi \omega_{-}}-\omega_{-}\cot{\pi \omega_{+}}}{2\omega_{+}\omega_{-}\qty(\omega_{+}^2-\omega_{-}^2)}+\pi\frac{\omega_{-}\cot{\pi \omega_{-}}-\omega_{+}\cot{\pi \omega_{+}}}{2\qty(\omega_{+}^2-\omega_{-}^2)}
\end{eqnarray*}
By collecting together all the factors calculated so far we get;
\begin{eqnarray*}
\mathcal{A}_x&=&\frac{\Pi_1}{2\pi}\int_{\mathbb{R}} \mathrm{d}k \,I(k)\,e^{-i\frac{D_x}{2\alpha} k^{2}}
=\Pi_1\sqrt{\frac{1}{1-2D_x A_y}}\,e^{iP_x}\\
\mathcal{A}_y&=&\Pi_1\sqrt{\frac{1}{1-2D_y A_x}}\,e^{iP_y}
\end{eqnarray*}
where the Phase $P_x \qty(P_y)$ depends on the electric field component $E_{x}\qty(E_y)$;
\begin{eqnarray*}
P_x&=&\frac{a^2}{\frac{D_x}{2\alpha}(8\Delta-4\lambda)^2+8\Delta-4\lambda}-\frac{a^2}{8\Delta-4\lambda}\\
P_y&=&\frac{b^2}{\frac{D_y}{2\alpha}(8\Delta+4\lambda)^2-8\Delta-4\lambda}+\frac{b^2}{8\Delta+4\lambda}
\end{eqnarray*}
\\
\\
Next, we rewrite the Amplitudes $\mathcal{A}_x, \mathcal{A}_y$ in terms of $\omega_x, \omega_y$ and $\Omega_+, \Omega_-$ by using the identities;
\begin{eqnarray*}
\omega_{+}\omega_{-}&=&\qty(\frac{T}{2\pi})^2\omega_{x}\omega_{y}\\
\omega_{+}^2-\omega_{-}^2&=&\qty(\frac{T}{2\pi})^2\,\Omega_{+}\Omega_{-}
\end{eqnarray*}
A simple but rather tedious calculation gives then;
\\
\\
\begin{eqnarray*}
\mathcal{A}_x&=&e^{iP_x}\sqrt{\frac{m\Omega_{+}\Omega_{-}}{2i\pi \hbar}}\sqrt{\frac{\omega_y}{\Omega_{-}(\omega_{x}+\omega_{y})\sin{\frac{\Omega_{+}T}{2}}-\Omega_{+}(\omega_{x}-\omega_{y})\sin{\frac{\Omega_{-}T}{2}}}}\\
\mathcal{A}_y&=&e^{iP_y}\sqrt{\frac{m\Omega_{+}\Omega_{-}}{2i\pi \hbar}}\sqrt{\frac{\omega_x}{\Omega_{-}(\omega_{x}+\omega_{y})\sin{\frac{\Omega_{+}T}{2}}+\Omega_{+}(\omega_{x}-\omega_{y})\sin{\frac{\Omega_{-}T}{2}}}}
\end{eqnarray*}
\\
\\
In conclusion, we obtain the total Amplitude $\mathcal{A}=\mathcal{A}_x\mathcal{A}_y$; 
\begin{equation}
\label{formula_amplitude}
\mathcal{A}=\frac{m\Omega_{+}\Omega_{-}e^{i P}}{2\pi i \hbar}\sqrt{\frac{\omega_{x}\,\omega_{y}}{(\omega_{x}+\omega_{y})^{2}\,\Omega_{-}^2\sin^{2}{\frac{\Omega_{+}T}{2}}-(\omega_{x}-\omega_{y})^{2}\,\Omega_{+}^2\sin^{2}{\frac{\Omega_{-}T}{2}}}}
\end{equation}
\\
where the Phase $P=P_x+P_y$ is;
\\
\begin{eqnarray*}
P=\frac{a^2}{4\alpha A_y}\qty(1-\frac{1}{1+2D_x})+\frac{b^2}{4\alpha A_x}\qty(1-\frac{1}{1+2D_y})
\end{eqnarray*}
The same result can be found in \cite{cervero:2017} where the Amplitude is calculated by means of the Stationary Phase Approximation.
\subsection{Limiting cases}
All the limiting cases can be derived directly from formula \ref{formula_amplitude}.\\  
\subsubsection{}
In the isotropic case $\omega_{x}=\omega_{y}=\omega$ and in the presence of a magnetic field $\vb{B}$ but vanishing electric field $\vb{E}$ the Amplitude reads as follows;
\\
\begin{equation}
\label{isotropic_oscillator}
\mathcal{A}=\frac{m\sqrt{\omega^2+\omega_{L}^2}}{2\pi i\hbar\sin\left(T\sqrt{\omega^2+\omega_{L}^2}\right)}
\end{equation}
Formula \ref{isotropic_oscillator} can be interpreted as the Amplitude of a two-dimensional \textit{isotropic} oscillator with effective frequency $\omega_{\text{eff}}=\sqrt{\omega^2+\omega_{L}^2}$ in both directions of the plane.
\\
\subsubsection{}
In the absence of any field ($\vb{E}=0, B=0$) formula \ref{formula_amplitude} reduces to the expression; 
\\
\begin{equation*}
\mathcal{A}=\sqrt{\frac{m \omega_{x}}{2\pi i\hbar\sin{\omega_{x}T}}}\cdot \sqrt{\frac{m \omega_{y}}{2\pi i\hbar\sin{\omega_{y}T}}}
\end{equation*}
\\
which is, as expected, the Amplitude of a two-dimensional anisotropic oscillator.\\
\section{\label{action}Classical trajectory and Action $S_{\text{cl}}$ }
In this section we determine the classical Action $S_{\text{cl}}$. In order to derive it we have to determine the classical trajectory by solving the classical equations of motion.\\
We consider the classical trajectory of a charged particle of charge $q$ and mass $m$ moving in static electric and magnetic fields. The fields are perpendicular to each other, \ie $\bold{B}=(0,0,B)$, $\bold{E}=(E_{x}, E_{y}, 0)$. In addition, there is a confining anisotropic oscillator potential. The particle's trajectory in the $(x,y)$-plane is determined by the following equations of motion;
\begin{align}
\begin{split}
\ddot{x}&=+\omega_{c}\dot{y}-\omega^{2}_{x}x +q E_{x}/m\\
\ddot{y}&=-\omega_{c}\dot{x}-\omega^{2}_{y}y+q E_{y}/m
\end{split}
\label{f11}
\end{align}
where $\omega_{c}=qB/m$ denotes the cyclotron frequency, whereas $\omega_{x}$ and $\omega_{y}$ are the oscillator frequencies in each direction of the 
plane.\\ 
\subsection{\label{traje}Classical Action for a charged Oscillator in a crossed electro- magnetic field.}
\label{caomf}
The effect of the electric field is trivial in the case of non-vanishing oscillators: $\omega_x\neq 0$, $\omega_y \neq 0$. It gives rise to a mere spatial shift by a fixed amount $\Delta x=qE_x/m\omega_x^2$ and $\Delta y=qE_y/m\omega_y^2$, respectively. We can therefore essentially ignore the electric field in this case.
By the following coordinate shift the electric field is eliminated in Eqs.~\ref{f11}. 
\begin{eqnarray*}
\bar{x}&=&x-\frac{qE_x}{m\omega_x^2}\\
\bar{y}&=&y-\frac{qE_y}{m\omega_y^2}\\
\end{eqnarray*}
The corresponding classical Action (an electric field included) can be retrieved from the zero-case $\vb{E}$ by substituting the variables $x_{0}, y_{0}, x_{1}, y_{1}$ with the shifted variables given by; 
\begin{equation}
\label{shifted_variables}
\begin{split}
\bar{x}_{0}&=x_0-\frac{qE_x}{m\omega_x^2}\\
\bar{x}_{1}&=x_1-\frac{qE_x}{m\omega_x^2}\\
\bar{y}_{0}&=y_0-\frac{qE_y}{m\omega_y^2}\\
\bar{y}_{1}&=y_1-\frac{qE_y}{m\omega_y^2}
\end{split}
\end{equation} 
We start by writing the system of Eqs.~\ref{f11} (setting $E_x=E_y=0$) in matrix form;
\\
\begin{equation}
\label{f22}
    \begin{pmatrix} \ddot{x} \\ \ddot{y} \end{pmatrix} =
    \underbrace{\begin{pmatrix} 0& \omega_{c} \\
        -\omega_{c} & 0 \end{pmatrix}}_{:=A}\begin{pmatrix} \dot{x} \\ \dot{y} \end{pmatrix}
\underbrace{-\begin{pmatrix} \omega^{2}_{x}& 0 \\
        0 & \omega^{2}_{y}  \end{pmatrix}}_{:=B}
\begin{pmatrix} x \\ y \end{pmatrix}
\end{equation}
\\
By introducing the two-dimensional vectors $\bold{X}:=(x, y)$ and $\bold{Y}:=\dot{\bold{X}}=(\dot{x}, \dot{y})$ we obtain a first order system;\\
\begin{equation}
\label{g2}
\begin{pmatrix} \dot{\bold{X}} \\ \dot{\bold{Y}} \end{pmatrix} =
    \underbrace{\begin{pmatrix} 0& \mathds{1}_{2} \\
        B & A \end{pmatrix}}_{:=M}\begin{pmatrix} \bold{X} \\ \bold{Y} \end{pmatrix}
\end{equation}
where the $4\cross 4$ matrix M is;
\begin{equation*}
M=\begin{pmatrix} 0& 0 & 1& 0 \\
        0 & 0 &0 & 1\\
 -\omega^{2}_{x} & 0 &0 & \omega_{c}\\
 0 & -\omega^{2}_{y} &-\omega_{c} & 0\end{pmatrix}
\end{equation*}
A solution of Eq.~\ref{g2} is given by;
\begin{equation*}
\begin{pmatrix} \bold{{X}} \\ \bold{{Y}}\end{pmatrix} =\sum^{4}_{i=1}a_{i}\,\mathrm{e}^{\lambda_{i}t}\bold{v}_{i}
\end{equation*}
Where $\lambda_{i}$ ($\bold{v}_{i}$) are the eigenvalues (eigenvectors) of the matrix M.\\
The characteristic polynomial for this matrix is;
\begin{equation*}
\label{f21}
\det(\lambda \mathds{1}_{4}-M)=\lambda^4+\lambda^{2}\qty(\omega^{2}_{c}+\omega^{2}_{x}+\omega^{2}_{y})+\omega^{2}_{x}\omega^{2}_{y}
\end{equation*}
The polynomial has the following four roots;
\begin{equation*}
\lambda_{1,2}=\pm i\, \overline{\omega}_{+},\quad
\lambda_{3,4}=\pm i\, \overline{\omega}_{-}
\end{equation*}
where
\begin{equation*}
\overline{\omega}_{\pm}:=\frac{2\pi}{T}\omega_{\pm}=\frac{\Omega_{+}\pm \Omega_{-}}{2}
\end{equation*}
The corresponding 4 eigenvectors $\bold{v_{i}}$ of $M$ are;
\begin{eqnarray*}
\bold{v_{1}}&=&\qty(1, i\, k_{+}/\overline{\omega}_{+}, -i\,\overline{\omega}_{+}, k_{+})\qquad \qty(\Leftrightarrow\lambda_{1}=-i\,\overline{\omega}_{+})\\
\bold{v_{2}}&=&\qty(1, -i\, k_{+}/\overline{\omega}_{+}, i\,\overline{\omega}_{+}, k_{+})\qquad \qty(\Leftrightarrow\lambda_{2}=i\,\overline{\omega}_{+})\\
\bold{v_{3}}&=&\qty(1, i\, k_{-}/\overline{\omega}_{-}, -i\,\overline{\omega}_{-}, k_{-})\qquad \qty(\Leftrightarrow\lambda_{3}=-i\,\overline{\omega}_{-})\\
\bold{v_{4}}&=&\qty(1, -i\, k_{-}/\overline{\omega}_{-}, i\,\overline{\omega}_{-}, k_{-})\qquad \qty(\Leftrightarrow\lambda_{4}=i\,\overline{\omega}_{-})
\end{eqnarray*}
where 
\begin{equation*}
k_{\pm}=\frac{\omega_{x}^2-\overline{\omega}_{\pm}^2}{\omega_{c}}
\end{equation*}
in explicit form;
\\
\begin{align*}
\begin{split}
x\qty(t)&=a_{1}e^{-i\overline{\omega}_{+}t}+a_{2}e^{i\overline{\omega}_{+}t}+a_{3}e^{-i\overline{\omega}_{-}t}+a_{4}e^{i\overline{\omega}_{-}t}\\
y\qty(t)&=i[\,a_{1}k_{+}/\overline{\omega}_{+}\,e^{-i\overline{\omega}_{+}t}-\,a_{2}k_{+}/\overline{\omega}_{+}\,e^{i\overline{\omega}_{+}t}+\,a_{3}k_{-}/\overline{\omega}_{-}\,e^{-i\overline{\omega}_{-}t}-\,a_{4}k_{-}/\overline{\omega}_{-}\,e^{i\overline{\omega}_{-}t}]
\end{split}\label{f12} 
\end{align*}
The coefficients $a_1, a_2, a_3, a_4$ are fixed by initial and final points;\\ 
\begin{equation*}
x(0)=x_0 \quad y(0)=y_0 \quad x(T)=x_1\quad y(T)=y_1
\end{equation*}
\\
It proves useful to rewrite the functions $x(t)$, $y(t)$ in the form;
\\
\begin{equation}
\label{uf}
\begin{split}
x\qty(t)=&A_1\cos{\overline{\omega}_{+}t}+iA_2\sin{\overline{\omega}_{+}t}+
A_3\cos{\overline{\omega}_{-}t}+iA_4\sin{\overline{\omega}_{-}}t\\
y\qty(t) =&A_1\frac{k_+}{\overline{\omega}_{+}}\sin{\overline{\omega}_{+}t}-
i\frac{k_+}{\overline{\omega}_{+}}A_2\cos{\overline{\omega}_{+}t}
+A_3\frac{k_{-}}{\overline{\omega}_{-}}\sin{\overline{\omega}_{-}t}-i\frac{k_-}{\overline{\omega}_{-}}A_4\cos{\overline{\omega}_{-}t}
\end{split}
\end{equation}
where
\begin{eqnarray*}
A_1&=&a_1+a_2\quad A_2=a_2-a_1\\
A_3&=&a_3+a_4\quad A_4=a_4-a_3
\end{eqnarray*}
The exakt expressions of the general coefficients $A_1, A_2, A_3, A_4$ are given in Appendix~\ref{general_case}. Note that the coefficients $A_{2}, A_{4}$ are imaginary so that the functions $x(t)$ and $y(t)$ are real valued.\\
\\
The coefficients $A_1, A_2, A_3, A_4$ are solutions of the linear system;
\begin{equation*}
L\bold{x}=\bold{c}:=(x_0, y_0, x_1, y_1)^t
\end{equation*}
where
\\
\begin{eqnarray*}
L&=&\begin{pmatrix} 1& 0& 1& 0 \\
0 & -i\frac{k_+}{\overline{\omega}_{+}}  &0&-i\frac{k_-}{\overline{\omega}_{-}} \\
\cos{\overline{\omega}_{+}T}& i\sin{\overline{\omega}_{+}T}&\cos{\overline{\omega}_{-}T} & i \sin{\overline{\omega}_{-}T} \\
 \frac{k_+}{\overline{\omega}_{+}}\sin{\overline{\omega}_{+}T} &  -i\frac{k_+}{\overline{\omega}_{+}}\cos{\overline{\omega}_{+}T}&  \frac{k_-}{\overline{\omega}_{-}}\sin{\overline{\omega}_{-}T} &  -i\frac{k_-}{\overline{\omega}_{-}}\cos{\overline{\omega}_{-}T}\end{pmatrix}
\end{eqnarray*}
With the coefficients given in Appendix~\ref{general_case} one can calculate the classical Action $S_{\text{cl}}$. With a Computer Algebra System such as MAPLE\texttrademark \, it is an easy task to calculate the integral;
\begin{equation}
S_{\text{cl}}=\int_{0}^{T} \mathcal{L}_{\text{cl}}\,d{t}
\end{equation}
\\
It does not make sense to reproduce the result here. 
\subsubsection{}
The coefficients $A_1, A_2, A_3, A_4$ for the \underline{isotropic} case $\omega_x=\omega_y=\omega$ are exposed in Appendix~\ref{isotropic_case}. A straightforward calculation of the classical Action gives (See also \cite{belardinelli:2019}); 
\\
\begin{equation}
\label{coc}
\begin{split}
{S}_{\text{cl}}&=\frac{m\omega_{\text{\text{eff}}}}{2\sin{{\omega_{\text{\text{eff}}}T}}}[\left({x}^2_{0}+{x}^2_{1}+{y}^2_{0}+{y}^2_{1}\right)\cos{\omega_{\text{\text{eff}}}T}\\&-2(x_0 x_1+y_0 y_1)\cos{\omega_{L}T}-2(y_0 x_1-x_0 y_1)\sin{\omega_{L}T}]
\end{split}
\end{equation}
where
\begin{eqnarray*}
\omega_{\text{eff}}:=\sqrt{\omega_{L}^2+\omega^2}
\end{eqnarray*}
In the presence of an electric field, the classical Action is given by formula~\ref{coc} where the variables $x_{0}, x_{1}, y_{0}, y_{1}$ are to be substituted by the shifted variables in Eqs. ~\ref{shifted_variables}.
\\
\subsection{Classical Action for a charged particle in crossed electric and magnetic fields}
In the case of vanishing oscillators $\omega_{x}=\omega_{y}=0$ the equations of motion Eqs.~\ref{f11} read;
\begin{align}
\label{g1}
\begin{split}
\ddot{x}&=+\omega_{c}\dot{y}+q E_{x}/m\\
\ddot{y}&=-\omega_{c}\dot{x}+q E_{y}/m
\end{split}
\end{align}
In terms of the complex variable $\gamma=x+iy$ we obtain the equation;
\begin{equation*}
\label{f9}
\ddot{\gamma}+i\,\omega_{c}\dot{\gamma}=q\epsilon/m
\end{equation*}
which has the general solution;
\begin{equation}
\label{f10}
\gamma(t)=K_{1}+K_{2}e^{-i\omega_{c}t} -it\epsilon/B         
\end{equation}
Eq.~\ref{f10} describes a moving particle whose trajectory is a superposition of a uniform circular motion with frequency $\omega_{c}$ and a rectilinear constant drift with speed $v=\sqrt{E^{2}_{x}+E^{2}_{y}}/B$.
The velocity $\bold{v}=(v_x, v_y)=\qty(E_{y}/B, -E_{x}/B, 0)$ is perpendicular to both the magnetic and the electric field vector. The general form is given by the well-known formula;
\begin{equation*}
\bold{v}=\frac{\bold{E}\cross \bold{B}}{B^2}
\end{equation*}
For the calculation of the classical action it is once again useful to eliminate the electric field in Eq.~\ref{g1}. This can be achieved by the following continuous coordinate shift;
\begin{equation}
\label{gra}
\bar{\gamma}:=\gamma+it\frac{\epsilon}{B}
\end{equation}
In terms of Cartesian coordinates this reads;
\begin{eqnarray*}
\bar{x}&=&x-\frac{E_y}{B}t=x-v_x t\\
\bar{y}&=&y+\frac{E_x}{B}t=y-v_y t
\end{eqnarray*}
We get then;
\begin{equation*}
\ddot{\bar{\gamma}}+i\,\omega_{c}\dot{\bar{\gamma}}=0
\end{equation*}
We already know the classical action $S_{\text{cl}}$ for a particle in a constant homogeneous magnetic field. In terms of shifted variables $\bar{x}, \bar{y}$ it reads;
\begin{equation*}
\bar{S}_{\text{cl}}^{B}=\frac{m\omega_{L}}{2\tan{\omega_{L}T}}\qty[(\bar{x}_1-\bar{x}_0)^2+(\bar{y}_1-\bar{y}_0)^2]+m\omega_{L}(\bar{x}_0 \bar{y}_1-\bar{x}_1\bar{y}_0)
\end{equation*}
Transforming back to unshifted variables by using;
\begin{eqnarray*}
\bar{x}_0&=&x_0 \\
\bar{y}_0&=&y_0 \\
\bar{x}_1&=&x_1-\frac{E_y}{B}T\\
\bar{y}_1&=&y_1+\frac{E_x}{B}T
\end{eqnarray*}
leads to;
\begin{equation}
\begin{split}
\label{this}
{S}_{\text{cl}}=&\frac{m\omega_{L}}{2\tan{\omega_{L}T}}\qty[\qty(x_1-\frac{E_y T}{B}-x_0)^2+\qty(y_1+\frac{E_x T}{B}-y_0)^2]\\
+&m\omega_{L}\qty(x_0 y_1-x_1 y_0 +x_0 \frac{E_x T}{B}+y_0 \frac{E_y T}{B})
\end{split}
\end{equation}
\section{Propagator for a charged particle in crossed $(\vb{E}, \vb{B})$-fields}
We dedicate a separate section to the case $\omega_{x}=\omega_{y}=0$ with non-vanishing electric and magnetic fields $\vb{E}\neq 0,B\neq 0$. One can derive the Amplitude directly from formula \ref{formula_amplitude}. The classical Action is given by Eq.~\ref{this}. This calculation is done in Appendix~\ref{last}.
However, it is instructive to derive the propagator also in a different way. Once again, we apply the transformation \ref{gra}. Under this change of variables the Lagrangian transforms as follows;
\begin{eqnarray*}
\bar{\mathcal{L}}&=&
\underbrace{\frac{m}{2}\dot{\bar{\gamma}}\dot{\bar{\gamma}}^{\ast}-\frac{im\omega_{L}}{2}(\bar{\gamma}^{\ast}\dot{\bar{\gamma}}-\bar{\gamma}\dot{\bar{\gamma}}^{\ast})}_{:=\bar{L}_{\epsilon=0}}\\
&+&\frac{m}{2}\frac{\abs{\epsilon}^2}{B^2}+\qty[i\frac{m}{2}\frac{\epsilon^{\ast}}{B}\dot{\bar{\gamma}}+\frac{q}{4}\epsilon^{\ast}\qty(t \dot{\bar{\gamma}}-\bar{\gamma})+\frac{q}{2}\epsilon^{\ast}\bar{\gamma}+\qcc*]
\end{eqnarray*}
By replacing the expression $t \dot{\bar{\gamma}}-\bar{\gamma}$ with $ \dv{t}(t\bar{\gamma}-2\bar{\gamma})$ we get the simplified expression;
\begin{eqnarray}
\label{alternative_derivation}
\bar{\mathcal{L}}=\bar{\mathcal{L}}_{\epsilon=0}+\frac{m}{2}\frac{\abs{\epsilon}^2}{B^2}
+\dv{t}F
\end{eqnarray}
where
\begin{equation*}
F=\qty(\frac{qt}{4}+\frac{i m}{2B})\epsilon^{\ast}\bar{\gamma}+\qcc*=\frac{m}{2B}(i+\omega_{L}t)\epsilon^{\ast}\bar{\gamma}+\qcc*
\end{equation*}
Remembering that two Lagrangians are equivalent (\ie they yield the same equations of motion, via the Euler-Lagrange equations) if their difference is a total time derivative. Therefore we conclude that 
the Lagrangian in Eq.~\ref{alternative_derivation} is equivalent to
\begin{eqnarray*}
\tilde{\mathcal{L}}=\bar{\mathcal{L}}_{\epsilon=0}+\frac{m}{2}\frac{\abs{\epsilon}^2}{B^2}
\end{eqnarray*}
The Propagator is;
\begin{eqnarray}
\label{yeah}
\bra{{\textbf{x}}_{f}, T}\ket{{\textbf{x}}_{i},0}= e^{\frac{i}{\hbar}\frac{\abs{\epsilon}^2}{B^2}T}\int\limits_{\gamma_{t}(0)}^{\gamma_{t}(T)}\mathcal{D}\bar{\gamma}\,e^{\frac{i}{\hbar}i \bar{S}_{\epsilon=0}}
=e^{\frac{i}{\hbar}\frac{\abs{\epsilon}^2}{B^2}T}e^{\frac{i}{\hbar}{S}_{\text{cl}}}\frac{m\omega_{L}}{2\pi i \hbar \sin{\omega_{L}T}}
\end{eqnarray}
Where ${S}_{\text{cl}}$ is given by Eq.~\ref{this}. 
\section{\label{qm}Energy eigenvalues}
\subsection{Energy spectrum}
In order to find the energy spectrum and the eigenstates of a charged anisotropic harmonic oscillator in static and crossed electro-magnetic fields one can solve the Eigenvalue equation $\mathcal{H}\Psi=E\Psi$ with Hamiltonian;
\begin{equation*}
\mathcal{H}=\frac{1}{2m}(\bold{p}-q\bold{A})^{2}+\frac{1}{2}m\omega^{2}_{x}x^2+\frac{1}{2}m\omega^{2}_{y}y^2-q\,E_{x}\cdot {x}-q\,E_{y}\cdot y
\end{equation*}\\
This calculation is not an entirely trivial task \cite{schuh:1985}.\\
We can however obtain the energy spectrum in a direct way. 
The classical trajectory of the particle is described by a superposition of two harmonic motions with respective frequencies $\bar{\omega}_{+},\bar{\omega}_{-}$ (See Eq.~\ref{uf}).\\ We can therefore conclude; 
\begin{equation*}
E(n,m)=\hbar\bar{\omega}_{+}\qty(n+\frac{1}{2})+\hbar\bar{\omega}_{-}\qty(m+\frac{1}{2})
\end{equation*}
where
\begin{equation*}
\bar{\omega}_{\pm}=\frac{\sqrt{\omega^{2}_{c}+\qty(\omega_{x}+\omega_{y})^2}
\pm\sqrt{\omega^{2}_{c}+\qty(\omega_{x}-\omega_{y})^2}}{2}
\end{equation*}
\newpage
\appendix
\section{Coefficients for the general case}
\label{general_case}
\begin{eqnarray*}
E&=&2D^2\qty(1-\cos{\Omega_{+}T})-2C^2\qty(1-\cos{\Omega_{-}T})\\
C&=&\frac{\Omega_{+}}{2\omega_{c}}\qty(\omega_{x}^2-\omega_{x}\omega_{y})\\
D&=&\frac{\Omega_{-}}{2\omega_{c}}\qty(\omega_{x}^2+\omega_{x}\omega_{y})\\
A_{1}E&=&\underbrace{\qty(D^2-C^2)}_{=\omega_{x}^3\omega_{y}}x_{0}+\qty(C+D)x_{0}\qty(C\cos{\Omega_{-}T}-D\cos{\Omega_{+}T})\\
&+&\underbrace{\frac{\Omega_{+}^2-\Omega_{-}^2}{4}}_{=\omega_{x}\omega_{y}}y_{0}\qty(C\sin{\Omega_{-}T}+D\sin{\Omega_{+}T})\\
&+&2\qty(C^2-D^2)x_{1}\sin{\frac{\Omega_{+}T}{2}}\sin{\frac{\Omega_{-}T}{2}}\\
&-&\frac{\Omega_{+}^2-\Omega_{-}^2}{2}y_{1}\qty(C\sin{\frac{\Omega_{-}T}{2}}\cos{\frac{\Omega_{+}T}{2}}+D\sin{\frac{\Omega_{+}T}{2}}\cos{\frac{\Omega_{-}T}{2}})\\
\\
-iA_{2}E&=&\qty(C+D)x_{0}\qty(D\sin{\Omega_{+}T}-C\sin{\Omega_{-}T})\\
&+&\frac{\Omega_{+}^2-\Omega_{-}^2}{4}y_{0}\qty(C\cos{\Omega_{-}T}+D\cos{\Omega_{+}T}-C-D)\\
&+&2\qty(CD+C^2)x_{1}\sin{\frac{\Omega_{-}T}{2}}\cos{\frac{\Omega_{+}T}{2}}\\
&-&2\qty(CD+D^2)x_{1}\sin{\frac{\Omega_{+}T}{2}}\cos{\frac{\Omega_{-}T}{2}}\\
&+&\qty(C+D)\frac{\Omega_{+}^2-\Omega_{-}^2}{2}y_{1}\sin{\frac{\Omega_{+}T}{2}}\sin{\frac{\Omega_{-}T}{2}}\\
\\
A_{3}E&=&(D^2-C^2)x_{0}+\qty(C-D)x_{0}\qty(C\cos{\Omega_{-}T}+D\cos{\Omega_{+}T})\\
&-&\frac{\Omega_{+}^2-\Omega_{-}^2}{4}y_{0}\qty(C\sin{\Omega_{-}T}+D\sin{\Omega_{+}T})\\
&-&2(C^2-D^2)x_{1}\sin{\frac{\Omega_{+}T}{2}}\sin{\frac{\Omega_{-}T}{2}}\\
&+&\frac{\Omega_{+}^2-\Omega_{-}^2}{2}y_{1}\qty(C\sin{\frac{\Omega_{-}T}{2}}\cos{\frac{\Omega_{+}T}{2}}+D\sin{\frac{\Omega_{+}T}{2}}\cos{\frac{\Omega_{-}T}{2}})\\
\\
-iA_{4}E&=&\qty(D^2+CD)x_{0}\sin{\Omega_{+}T}+\qty(C^2-CD)x_{0}\sin{\Omega_{-}T}\\
&+&\frac{\Omega_{+}^2-\Omega_{-}^2}{4}y_{0}\qty(C\cos{\Omega_{-}T}-D\cos{\Omega_{+}T}+D-C)\\
&+&2\qty(CD-D^2)x_{1}\sin{\frac{\Omega_{+}T}{2}}\cos{\frac{\Omega_{-}T}{2}}\\
&+&2\qty(CD-C^2)x_{1}\sin{\frac{\Omega_{-}T}{2}}\cos{\frac{\Omega_{+}T}{2}}\\
&+&\qty(D-C)y_{1}\frac{\Omega_{+}^2-\Omega_{-}^2}{2}\sin{\frac{\Omega_{+}T}{2}}\sin{\frac{\Omega_{-}T}{2}}
\end{eqnarray*}
\section{Coefficients for the isotropic case $\bold{\omega_{x}=\omega_{y}}$}
\label{isotropic_case}
In the case of an isotropic harmonic potential $\omega_{x}=\omega_{y}=\omega$ we get the following coefficients;
\begin{eqnarray*}
\Omega&=&\sqrt{\omega_{c}^2+4\omega^2}=2\sqrt{\omega_{L}^2+\omega^2}=2\omega_{\text{eff}}\\
E&=&2\omega^4\qty(1-\cos{\Omega T})\\
A_{1}E&=&\omega^{4}x_{0}\qty(1-\cos{\Omega T})+\omega^{4}y_{0}\sin{\Omega T}\\
&-&2\omega^{4}x_{1}\sin{{\omega_{\text{eff}} T}}\sin{{\omega_{L}T}}\\
&-&2\omega^{4}y_{1}\sin{{\omega_{\text{eff}} T}}\cos{{\omega_{L}T}}\\
\\
-iA_{2}E&=&\omega^{4}x_{0}\sin{\Omega T}-\omega^{4}y_{0}\qty(1-\cos{\Omega T})\\
&-&2\omega^{4}x_{1}\sin{{\omega_{\text{eff}} T}}\cos{{\omega_{L}T}}\\
&+&2\omega^{4}y_{1}\sin{{\omega_{\text{eff}} T}}\sin{{\omega_{L}T}}\\
\\
A_{3}E&=&\omega^{4}x_{0}\qty(1-\cos{\Omega T})-\omega^{4}y_{0}\sin{\Omega T}\\
&+&2\omega^{4}x_{1}\sin{{\omega_{\text{eff}} T}}\sin{{\omega_{L}T}}\\
&+&2\omega^{4}y_{1}\sin{{\omega_{\text{eff}} T}}\cos{{\omega_{L}T}}\\
\\
-iA_{4}E&=&\omega^{4}x_{0}\sin{\Omega T}+\omega^{4}y_{0}\qty(1-\cos{\Omega T})\\
&-&2\omega^{4}x_{1}\sin{{\omega_{\text{eff}} T}}\cos{{\omega_{L}T}}\\
&+&2\omega^{4}y_{1}\sin{{\omega_{\text{eff}} T}}\sin{{\omega_{L}T}}\\
\end{eqnarray*}
By applying elementary trigonometric identities we can simplify the above formulae;
\\
\begin{eqnarray*}
A_{1}&=&\frac{1}{2}\qty(x_0+y_0\cot{{\omega_{\text{eff}} T}}-\frac{x_1\sin{{\omega_{L} T}}+y_1\cos{{\omega_{L} T}}}{\sin{{\omega_{\text{eff}} T}}})\\
\\
A_{2}&=&\frac{i}{2}\qty(-y_0+x_0\cot{{\omega_{\text{eff}} T}}-\frac{x_1\cos{{\omega_{L} T}}-y_1\sin{{\omega_{L} T}}}{\sin{{\omega_{\text{eff}} T}}})\\
\\
A_{3}&=&\frac{1}{2}\qty(x_0-y_0\cot{{\omega_{\text{eff}} T}}+\frac{x_1\sin{{\omega_{L} T}}+y_1\cos{{\omega_{L} T}}}{\sin{{\omega_{\text{eff}} T}}})\\
\\
A_{4}&=&\frac{i}{2}\qty(y_0+x_0\cot{{\omega_{\text{eff}} T}}-\frac{x_1\cos{{\omega_{L} T}}-y_1\sin{{\omega_{L} T}}}{\sin{{\omega_{\text{eff}} T}}})\\
\end{eqnarray*}
\section{Propagator for a charged particle in crossed $(\vb{E}, \vb{B})$-fields}
\label{last}
We start with the expression;
\begin{eqnarray*}
\mathcal{A}_x&=&\frac{\Pi_1}{2\pi}\int_{\mathbb{R}} \mathrm{d}k \int_{\mathbb{R}} \mathrm{d}x_0 \,e^{i(a+k)x_0}\,e^{-i\frac{D_x}{2\alpha} k^{2}}
\end{eqnarray*}
We can now use;
\begin{eqnarray*}
\int_{\mathbb{R}} \mathrm{d}x_0 \,e^{i(a+k)x_0}=2\pi\,\delta(a+k)
\end{eqnarray*}
resulting in:
\begin{eqnarray*}
\mathcal{A}_x=\Pi_1\int_{\mathbb{R}} \mathrm{d}k\, \delta(a+k)e^{-i\frac{D_x}{2\alpha} k^{2}}
=\Pi_1\,e^{-i\frac{D_x}{2\alpha}a^2}
\end{eqnarray*}
where 
\begin{eqnarray*}
\Pi_1=\lim_{\omega \to 0}\frac{1}{2}\sqrt{\frac{\alpha \omega_{+}\omega_{-}}{i\pi \sin{\pi \omega_{+}}\sin{\pi \omega_{-}}}}=\sqrt{\frac{m\omega_{L}}{2\pi i\hbar\sin{\omega_{L}T}}}
\end{eqnarray*}
In addition,
\begin{eqnarray*}
\lim_{\omega\to 0}D_x=\sum_{n=1}^{\infty}\frac{1}{n^2-B_0}=\frac{1}{2B_0}-\frac{\pi}{2\sqrt{B_0}}\cot{\pi \sqrt{B_0}}
\end{eqnarray*}
By using
$B_0=\lim_{\omega\to 0}B=\qty(\frac{T}{2\pi})^{2}\omega_{c}^2$
we obtain
\begin{eqnarray*}
\lim_{\omega\to 0}D_x=\frac{1}{2}\qty(\frac{\pi}{\omega_{L}T})^2\qty(1-\omega_{L}T\cot{\omega_{L}T})
\end{eqnarray*}
Altogether, this leads to the Amplitudes;
\begin{eqnarray*}
\mathcal{A}_x&=&\sqrt{\frac{m\omega_{L}}{2\pi i\hbar\sin{\omega_{L}T}}}e^{-i\frac{mE_{x}^2}{2 B^2}\frac{T}{\hbar}}e^{i\frac{m\omega_{L}}{2\hbar}\frac{E_{x}^2 T^2}{B^2}\cot{\omega_{L}T}}\\
\mathcal{A}_y&=&\sqrt{\frac{m\omega_{L}}{2\pi i\hbar\sin{\omega_{L}T}}}e^{-i\frac{mE_{y}^2}{2 B^2}\frac{T}{\hbar}}e^{i\frac{m\omega_{L}}{2\hbar}\frac{E_{y}^2 T^2}{B^2}\cot{\omega_{L}T}}
\end{eqnarray*}
The full Propagator is then given by;
\begin{eqnarray}\label{yeah2}
K=e^{\frac{i}{\hbar} S_{\text{cl}}}\mathcal{A}_x\mathcal{A}_y
=e^{\frac{i}{\hbar} S_{\text{cl}}}\frac{m\omega_{L}}{2\pi i\hbar\sin{\omega_{L}T}}e^{-i\frac{m}{2\hbar}\frac{\abs{\bold{E}}^2}{B^2}T}e^{i\frac{m\omega_{L}}{2\hbar}\frac{\abs{\bold{E}}^2 T^2}{B^2}\cot{\omega_{L}T}}
\end{eqnarray}
where
\begin{equation*}
\begin{split}
{S}_{\text{cl}}=&\frac{m\omega_{L}}{2\tan{\omega_{L}T}}\qty[\qty(x_1-\frac{E_y T}{B}-x_0)^2+\qty(y_1+\frac{E_x T}{B}-y_0)^2]\\
+&m\omega_{L}\qty(x_0 y_1-x_1 y_0 +x_0 \frac{E_x T}{B}+y_0 \frac{E_y T}{B})
\end{split}
\end{equation*}
The Propagator in Eq.~\ref{yeah2} is equivalent to the one in Eq.~\ref{yeah}.
\bibliography{Propagator_HO_static_electromagnetic_field}
\bibliographystyle{plain}
\end{document}